\documentclass[a4paper]{jpconf}
\usepackage{graphicx}
\usepackage{amsmath}
\usepackage{amssymb}
\usepackage{caption}
\usepackage{subcaption}
\usepackage{multicol}
\usepackage{cite}
\usepackage[bookmarks=true,
   colorlinks=true,
   linkcolor=blue,
   urlcolor=blue,
   citecolor=blue,
   bookmarks=true,
   hyperindex=true
]{hyperref}
\usepackage[dvipsnames]{xcolor}

\begin{document}
\title{Quantum and quantum-inspired optimization for an in-core fuel management problem}

\author{S~R~Usmanov$^{*}$, G~V~Salakhov, A~A~Bozhedarov, E~O~Kiktenko and A~K~Fedorov$^{\dagger}$}

\address{Russian Quantum Center, Skolkovo, Moscow 121205, Russia}

\ead{$^{*}$s.usmanov@rqc.ru; $^{\dagger}$akf@rqc.ru}

\begin{abstract}
Operation management of nuclear power plants consists of several computationally hard problems. 
Searching for an in-core fuel loading pattern is among them.
The main challenge of this combinatorial optimization problem is the exponential growth of the search space with a number of loading elements. 
Here we study a reloading problem in a Quadratic Unconstrained Binary Optimization (QUBO) form. 
Such a form allows us to apply various techniques, including quantum annealing, classical simulated annealing, and quantum-inspired algorithm in order to find fuel reloading patterns for several realistic configurations of nuclear reactors. 
We present the results of benchmarking the in-core fuel management problem in the QUBO form using the aforementioned computational techniques. 
This work demonstrates potential applications of quantum computers and quantum-inspired algorithms in the energy industry.
\end{abstract}

\section{Introduction}

Quantum computing has emerged as a promising paradigm for tackling complex computational challenges in various fields, including industrial applications~\cite{Yarkoni2022,Fedorov2021,Boev2023, Fedorov2022, Orus2019}. 
As it is expected, the intrinsic quantum properties, such as superposition and entanglement, enable quantum computers to efficiently solve certain problems that are intractable for classical computers. 
In this context, quantum annealing has shown significant potential for addressing discrete optimization problems, where the objective is to find the best vector in a very large solution space.

One industrial problem that demands efficient optimization techniques is the in-core reloading patterns optimization in nuclear reactors~\cite{DoBinh2014,Meneses2018ArtificialIM}. 
In-core reloading is a critical process in nuclear power plants, involving the arrangement of fuel assemblies within the reactor core to ensure optimal power distribution and safe operation. 
This process is routinely performed during planned reactor shutdowns, which can occur on the order of months or years depending on the reactor type and operational schedule.
The main goal of in-core reloading is to maximize the reactor's power output while adhering to stringent safety constraints, such as maintaining sub-criticality and minimizing the likelihood of local power peaking.

Conventional methods for searching reloading patterns in the local space, which are typically limited to about one-fourth of the reactor core, rely on heuristic algorithms~\cite{LI2022111950}. 
These heuristics are effective to a certain extent, but they do not guarantee to find the globally optimal solution. 
Furthermore, as the complexity and size of nuclear cores increase, these classical approaches encounter limitations in providing the best possible reloading patterns in a reasonable amount of time. 
While there have been previous attempts to address the in-core reloading problem with quantum computing techniques, some crucial elements of the solution, such as objective function and constraints, have not been defined~\cite{Whyte2020}. 

In the present paper, we present an approach to the in-core reloading patterns optimization problem by leveraging quantum annealing with the QUBO formulation using additional constraints for the symmetrical loading of fuel elements in the core. 
The proposed QUBO model is designed to encode the problem's constraints, transforming the complex optimization problem into a suitable form for quantum annealers. 
We demonstrate the proof-of-concept by testing the proposed method on existing and synthetic nuclear cores.
In summary, this paper aims to contribute to the growing knowledge on the application of quantum computing for industrial problem-solving by offering a novel QUBO-based method for in-core reloading patterns optimization in nuclear reactors. 
Through this research, we hope to pave the way for more efficient and reliable solutions to the crucial problem in the nuclear industry, with potential implications for the safe and optimized operation of nuclear power plants.

\section{In-core reloading pattern heuristics}

The best possible solution for finding reloading patterns requires complex physical modeling of the nuclear core. 
Ref.~\cite{Galperin1999} has suggested several forbiddance rules that in fact convert reloading optimization into a combinatorial constraint satisfaction problem.
The cycle number (burnup state) of the fuel is equated to enrichment in order to apply this rule set to the main design tasks in this paper. 
As a result, the fuel with the highest enrichment is equivalent to fresh fuel in a batch system, and the fuel with the lowest enrichment is equivalent to fuel that has been twice burned.

Taking into account the considerations from Refs.~\cite{Whyte2020, Horelik2013}, we arrive to the following rules:
\begin{enumerate}
	\item a fresh element should not be placed into innermost core positions;
	\item twice-burned fuel core element should not be placed into the peripheral core position;
	\item a pair of fresh fuel elements should not be placed in adjacent positions unless one of them is adjacent to a peripheral one;
	\item a pair of twice-burned fuel elements should not be placed in adjacent positions.
\end{enumerate}

\begin{figure}[!h]
\centering
\includegraphics[width=0.4\linewidth]{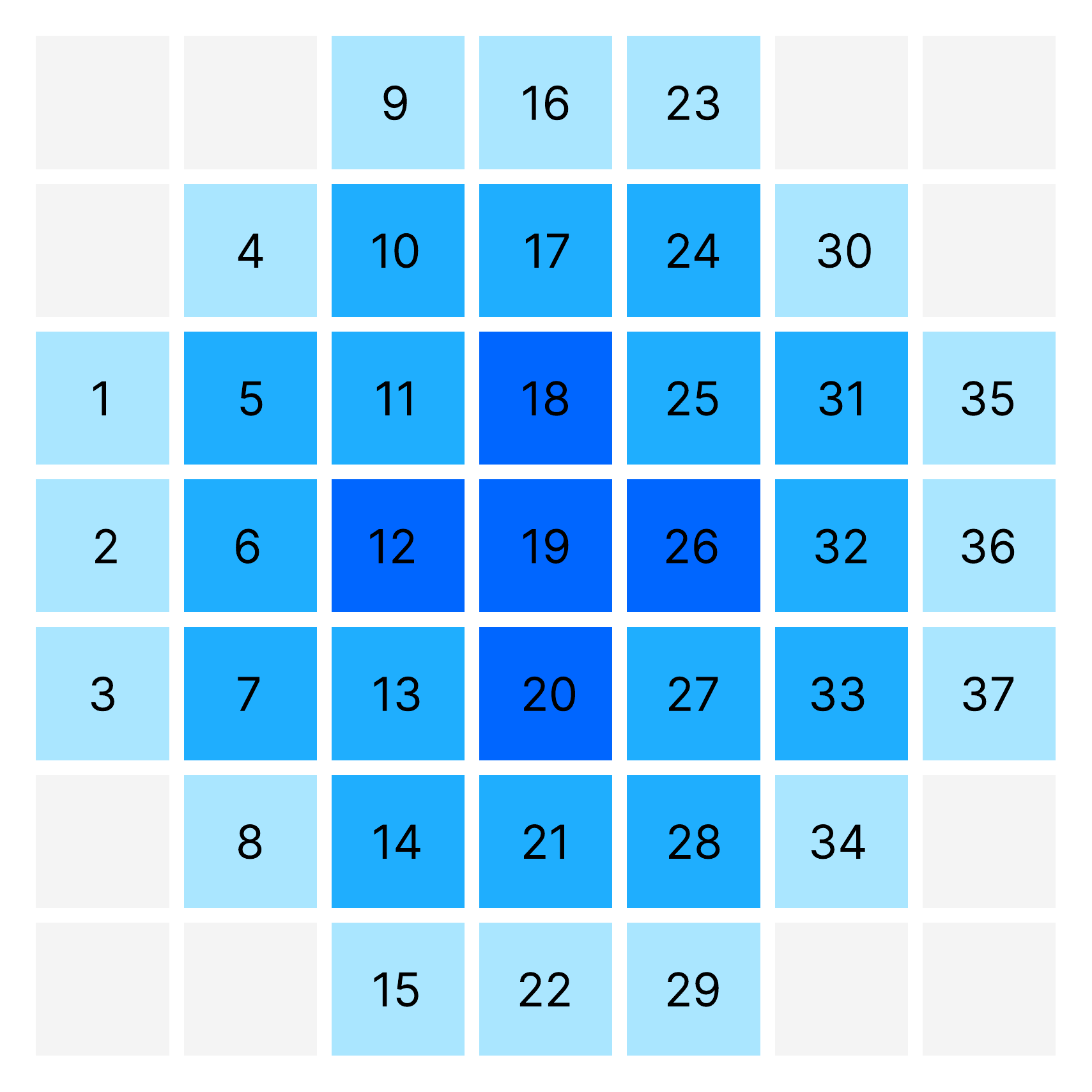}
\caption{In-core cells numeration. Light color corresponds to peripheral positions, blue color corresponds to middle positions and dark blue color denotes inner core positions.}
\label{fig:cells_numeration}
\end{figure}

An example of the core structure is shown in Fig.~\ref{fig:cells_numeration}.

To formulate the above forbiddance rules as a mathematical problem, we introduce binary decision variables $x_i^b$, that indicate which of $N$ fuel assemblies should be placed at position $i\in\{1,2,\ldots,N\}=:{\cal N}$, index $b \in \{0,1,2\}$ 
corresponds to fresh, once-burned, and twice-burned fuel elements, respectively (we are considering 3 cycles for fuel elements case):
\begin{equation}
	x^{0}_{i} =
	\begin{cases}
		1, & \text{if fresh fuel assembly $i$ is in the core position $i$}; \\
		0, & \text{otherwise};
	\end{cases}
\end{equation}
\begin{equation}
	x^{1}_{i} =  \begin{cases}
		1, & \text{if once-burnt fuel assembly $i$ is in the core position $i$}; \\
		0, & \text{otherwise};
\end{cases}
\end{equation}
\begin{equation}
	x^{2}_{i} =
	\begin{cases}
		1, & \text{if twice-burnt fuel assembly $i$ is in the core position $i$}; \\
		0, & \text{otherwise}.
	\end{cases}
\end{equation}
We denote border (peripheral) and innermost core positions subsets as $\mathcal{B}\subset {\cal N}$ and $\mathcal{I} \subset {\cal N}$ correspondingly ($\mathcal{B} \cap \mathcal{I}=\emptyset$).
We denote $N^{(b)}$ with $b\in\{0,1,2\}$ the total number of fresh, once-burned, and twice-burned fuel elements correspondingly ($\sum_{b=0}^2 N^{b}=N$).

Then forbiddance rules can be formulated as a constraint satisfaction problem. 
Cell loading constraint sets exactly one fuel element to every cell in the core:
\begin{equation}\label{constr1}
	x^{0}_{i}+x^{1}_{i}+x^{2}_{i}=1, \quad \forall i \in {\cal N}.
\end{equation}
The final number of assemblies of each type constrains:
\begin{equation}\label{quant_constr}
\sum\limits_{i\in\mathcal{N}}^{}x^{b}_{i}=N^{b},\quad b\in\{0,1,2\}.
\end{equation}

Border core positions should not be loaded by twice burnt fuel assemblies according to rule (i) constraint:
\begin{equation}\label{constr2}
\sum\limits_{i\in\mathcal{B}}^{}x^{2}_{i}=0.
\end{equation}
Inner core positions should not be loaded by fresh fuel assemblies according to rule (ii) constraint:
\begin{equation}\label{constr3}
	\sum\limits_{i\in\,\mathcal{I}}^{}x^{0}_{i}=0.
\end{equation}

\begin{figure}[!h]
\centering
\includegraphics[width=0.9\linewidth]{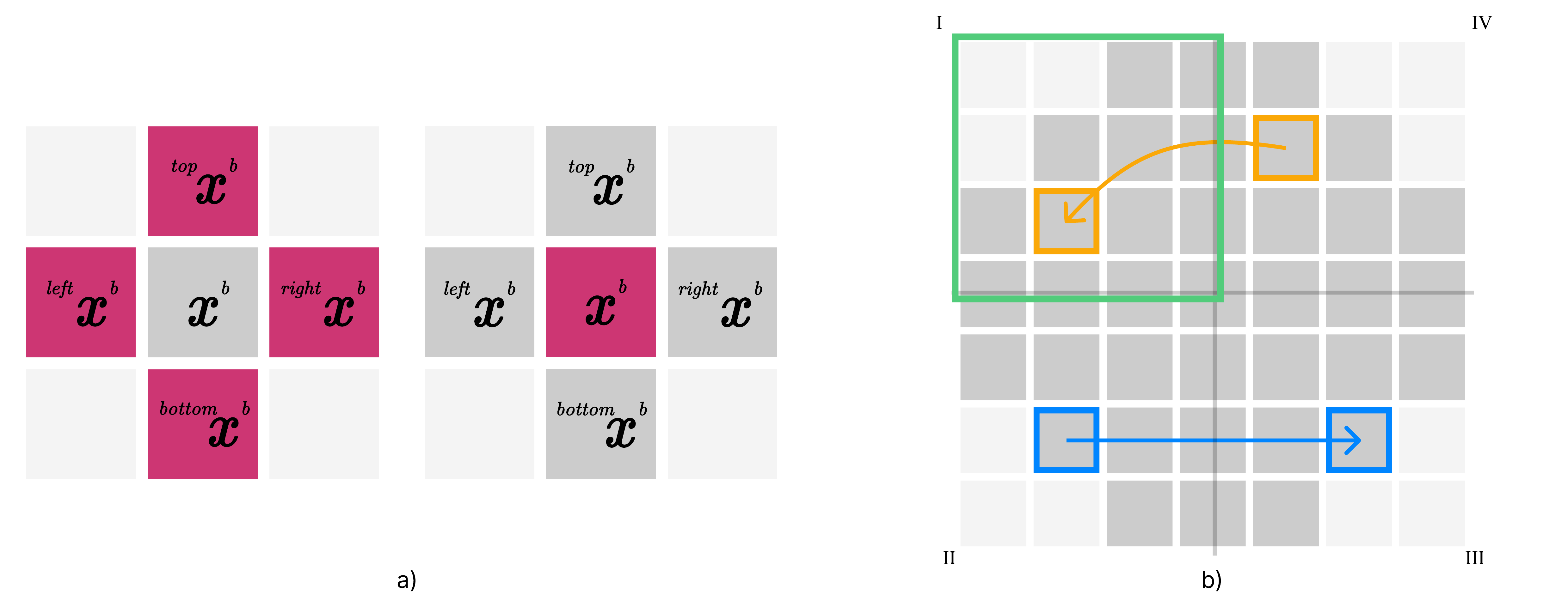}
\caption{a) Example of valid configurations for constraints given by Eq.~(\ref{rule3}) and Eq.~(\ref{rule4}). b) Rotational symmetry (yellow) maps the cell from the 4th to 1st quadrant and horizontal mirror symmetry (blue) maps the core cell from the 2nd to 3rd quadrant.}
\label{fig:symmetry_example}
\end{figure}

Fresh fuel elements adjacent positions constraint according to  rule (iii):
\begin{equation}\label{rule3}
    \begin{aligned}
        x^{0}_{i} +  ^{\text{left}}x^{0}_{i} \leq 1\text{~if~}{\rm left}(i)\not\in{\cal B}, &\quad
        x^{0}_{i} +  ^{\text{right}}x^{0}_{i} \leq 1\text{~if~}{\rm right}(i)\not\in{\cal B}\\
        x^{0}_{i} +  ^{\text{top}}x^{0}_{i} \leq 1\text{~if~}{\rm top}(i)\not\in{\cal B}, &\quad
        x^{0}_{i} +  ^{\text{bottom}}x^{0}_{i} \leq 1\text{~if~}{\rm bottom}(i)\not\in{\cal B}\\
    &\forall i\in{\cal N} \setminus {\cal B},
    \end{aligned}
\end{equation}
where $^{\rm dir}x_i^b$ denotes the value of $x_j^b$ located in the neighboring position $j$ in the direction ${\rm dir}$ with respect to $i$, or is 0 if there is no such position $j$, and ${\rm dir}(i)$ is a corresponding neighboring position [see Fig.~\ref{fig:symmetry_example}(a)].

Twice-burnt fuel elements adjacent positions constraint according to rule (iv):
\begin{equation}\label{rule4}
    \begin{aligned}
        x^{2}_{i} +  ^{\text{left}}x^{2}_{i} \leq 1, &\quad
        x^{2}_{i} +  ^{\text{right}}x^{2}_{i} \leq 1\\
        x^{0}_{i} +  ^{\text{top}}x^{2}_{i} \leq 1, &\quad
        x^{0}_{i} +  ^{\text{bottom}}x^{2}_{i} \leq 1\\
    &\forall i\in{\cal N}.
    \end{aligned}
\end{equation}

By mapping the in-core reloading problem to a QUBO form, we encode the problem's constraints, objective function, and decision variables in a way that quantum annealers can efficiently explore the solution space. 

Since the resulting configuration usually requires additional rotational or mirror symmetry~\cite{Lindley2014, Meneses2010, Dehart1995}, extra constraints could be added in the following way:
\begin{eqnarray}
    &x^{b}_{i} = \: ^{\text{rot}(\pi/2)}x^{b}_{i} = \: ^{\text{rot}(\pi)}x^{b}_{i} = \:^{\text{rot}(3\pi/2)}x^{b}_{i}, \quad &\forall i \in \mathcal{Q}_{\rm I}, b \in \{0,1,2\}, \label{constr6} \\
    &x^{b}_{i} = \: ^{\text{ref}(v)}x^{b}_{i} = \: ^{\text{ref}(h)}x^{b}_{i} = \: ^{\text{ref}(v, h)}x^{b}_{i}, \quad &\forall i \in \mathcal{Q}_{\rm I}, b \in \{0,1,2\},\label{constr7}
\end{eqnarray}
where $^{\text{rot}(\theta)}x^{b}_{i}$ is the value of $x^{b}_{j}$ in the position $j$ obtained from $i$ by counterclockwise rotation on $\theta$, similarly, $\text{ref}(axis)$ denotes a reflection around one or two axes ($axis=v$ is vertical, $axis=h$ is horizontal, and $\text{ref}(h,v)$ means both axis simultaneously),
and ${\cal Q}_{\rm I}\subset{\cal N}$ denotes a subset of positions in the upper left corner [see Fig.~\ref{fig:symmetry_example}(b)].
Rotational and mirror symmetry examples are shown in Fig.~\ref{fig:symmetry_example}. 

\section{QUBO formulation of in-core reloading pattern heuristics}

To tackle the in-core reloading problem using quantum annealing, it is required to transform it into a suitable QUBO model. 
The QUBO formulation provides a way to represent combinatorial optimization problems in a form compatible with quantum annealing devices such as quantum annealers.
The QUBO model is a well-established formalism used to represent combinatorial optimization problems as quadratic polynomials over binary variables and may be formulated in the following way: 
\begin{equation}
	H =  \boldsymbol z^T Q \boldsymbol z \to \min.
\end{equation}
where $\boldsymbol z=(\boldsymbol x^0, \boldsymbol x^1, \boldsymbol x^2)$ is the vector of binary variables that concatenate sets of burn level variables, and $Q$ is a real-valued upper triangular or symmetric matrix. 

Below we show a transformation of constraints given by Eqs.~(\ref{constr1})-(\ref{constr7}) to the QUBO form. 
Constraint~(\ref{constr1}) can be reformulated as:
\begin{equation}\label{h-first}
  H^{(1)}  = \sum\limits_{i = 1 }^{N}\left(x^{0}_{i}+x^{1}_{i}+x^{2}_{i} - 1\right)^2.
\end{equation}

Quantity constraints (\ref{quant_constr}) take the form:
\begin{equation}\label{H_constr2}
 H^{(2)} = \sum_{b=0}^2\left(\sum\limits_{i\in\mathcal{N}}^{}x^{b}_{i}-N^{b}\right)^2.
\end{equation}

Constraints (\ref{constr2}) and (\ref{constr3}) can be reformulated as:
\begin{equation}\label{qubo-constr2-3}
 H^{(3)} = \left(\sum\limits_{i\in\,\mathcal{B}}^{}x^{1}_{i}\right)^2 + \left(\sum\limits_{i\in\,\mathcal{I}}^{}x^{0}_{i}\right)^2.
\end{equation}

Constraints inequalities (\ref{rule3})-(\ref{rule4}) could be represented \cite{Glover2019, Boev2023} in quadratic terms. 
We suggest using a precalculated number of adjacent elements not on the border:
\begin{equation}
	H^{(4)} = \sum\limits_{i\in{\cal N}}\sum\limits_{\substack{k \in {\rm nb}(i),\\ k \notin \mathcal{B}}}^{} x^0_{i} x^0_{k} + \sum\limits_{i\in{\cal N}}\sum\limits_{k \in  {\rm nb}(i)}^{} x^2_{i} x^2_{k},
\end{equation}
where ${\rm nb}(i)$ denotes neighboring positions adjacent to $i$.

Symmetry constraints (\ref{constr6}) and (\ref{constr7}) could be added as the sum of coupling with the left upper quadrant:
\begin{equation}\label{h-last}
	H^{(5)} = \sum\limits_{b=1}^{3}\sum\limits_{i \in \mathcal{Q}_{\rm I}}\sum\limits_{k \in \mathcal{S}_i^1}^{} (x^b_{i} - x^b_{k})^2 + \sum\limits_{b=1}^{3}\sum\limits_{i \in \mathcal{Q}_{\rm I}}\sum\limits_{k \in \mathcal{S}_i^2} (x^b_{i}-x^b_{k})^2,
\end{equation}
where $\mathcal{S}_i^1$ and $\mathcal{S}_i^2$ denote rotational and mirror symmetry core positions sets of element $i$, correspondingly.
The resulting expression for $H$ is obtained by sum of Eqs.(\ref{h-first})--(\ref{h-last}):
\begin{equation}
	H = H^{(1)} + H^{(2)} + H^{(3)} + H^{(4)} + H^{(5)}.
\end{equation}

\section{Quantum and quantum-inspired annealing}

Quantum annealing is a specialized quantum computing technique designed to solve combinatorial optimization problems efficiently. 
At its core, quantum annealing exploits the principles of quantum mechanics, specifically quantum tunneling and fluctuations, to explore the solution space of an optimization problem and find the lowest-energy configuration or ground state. 
The quantum annealing process can be seen as a quantum analog of simulated annealing, a classical optimization algorithm that emulates the annealing process in metallurgy.

The quantum annealing technique relies on the optimization problem being encoded into the Hamiltonian $H$, representing the energy of the system, which typically consists of two components: 
the objective function $H_O$ (encoding the problem's constraints and goals) and a ``driver" Hamiltonian $H_D$ (encouraging exploration). 
The quantum annealing device, such as a quantum annealer, then evolves the system over time $t$, transitioning from a quantum superposition state to the ground state that encodes the optimal solution:
\begin{equation}
	H(t) = \left(1-\frac{t}{T}\right)H_D + t H_O, \quad t\in[0; T],
\end{equation}
where $T$ is the annealing time.

In practice, quantum annealers are designed to operate on Ising models, which represent problems using spin variables. 
QUBO model could be transformed into the Ising model by variables mapping:
\begin{equation}
	s_i = 2 z_i - 1 \quad s_i \in \{-1; 1\}.
\end{equation}
This conversion process ensures that the benefits of quantum annealing can be effectively harnessed for solving optimization problems that are initially formulated in the QUBO form.

The Coherent Ising Machine (CIM)~\cite{Yamamoto2017} is a specialized quantum-inspired annealer that emulates Ising model-based optimization problems using coherent quantum operations. 
CIM devices use laser pulses to encode the Ising problem into the quantum states of photons, which then undergo coherent interactions to explore the solution space. 
These machines aim to achieve computational speedup while still operating in a coherent, classical regime.
Due to the limited availability and practical constraints of CIM hardware, researchers have developed software emulators of CIMs. 
Such software simulates the behavior of coherent Ising machines on classical computers, providing a means to test and validate quantum-inspired algorithms and QUBO-based formulations without access to dedicated quantum hardware.

In the context of the in-core reloading patterns optimization problem, we explore the potential of quantum annealing with the QUBO formulation. 
Additionally, we consider the software emulator of a coherent Ising machine, SimCIM, as an alternative quantum-inspired approach.

\section{Benchmarking}

Our methodology encompasses extensive experimentation and benchmarking of quantum annealer D-Wave, quantum-inspired algorithm SimCIM (see Ref.~\cite{Tiunov2019}), and a classical QUBO solver. 
By presenting the results of real nuclear core tests, we aim to validate the effectiveness and scalability of our method, which paves the way for the wider adoption of quantum computing in solving complex optimization challenges in the nuclear industry and beyond.

To validate the QUBO form of the reloading pattern problem, we searched for several reactor configurations taking into account constraints and different types of symmetry. 
For details see Appendix (Fig.~\ref{fig:symms}).
We run the algorithm on seven different configurations of nuclear cores. 
The number of core cells and an available number of fresh, once burnt, and twice burnt fuels elements, are represented in Table~\ref{tab:tts-conf}.
We have used three real nuclear core configurations such as Angra nuclear core~\cite{Sabundjian2012}, a 4-loop PWR design by Westinghouse~\cite{Freixa2007}, and NuScale's Small Modular Reactor~\cite{INGERSOLL201484}. 
The rest of the tested cores are synthetic. 

\begin{table}[!h]
    \caption{Number of fuel elements for different reactor cores (see Figure~\ref{fig:feasible_tts}).}
    \label{tab:tts-conf}
	\centering
	\begin{tabular}{llll}
	\br
	Cells number $N$ & Fresh $N^0$ & Once burnt $N^1$ & Twice burnt $N^2$ \\ \mr
	13           & 8     & 4           & 1           \\ 
	37           & 16    & 17          & 4           \\ 
	69           & 33    & 21          & 15          \\ 
	97           & 36    & 36          & 25          \\
	121          & 49    & 45          & 36          \\ 
	193          & 64    & 80          & 49          \\ 
	241          & 75    & 100         & 65          \\ \br
	\end{tabular}
\end{table}

The core configuration and the number of fuel elements have been used to create QUBO form of the loading pattern problem, and then QUBO form has been transformed into Ising model. 
The corresponding Ising problems have been solved by real quantum devices and SimCIM emulators of quantum computers. 
A simulated annealing algorithm has been chosen as a classical solver. 

We have used Time-to-Solution (TTS) metric~\cite{Aramon2019, Stollenwerk2020} to compare different approaches in solving QUBO form of reloading pattern problem. 
\begin{equation}
	\rm{TTS} = \tau_a R_{99}.
\end{equation}
Here $\tau_a$ is a single annealing time and $R_{99}$ is the number of required repetitions to get a solution with the probability 0.99:
\begin{equation}
    R_{99} = \frac{\log(1 - 0.99)}{\log(1 - \theta)}, 
\end{equation}
where $\theta$ is success probability.

\begin{figure}[!h]
    \centering
    \includegraphics[width=1.0\linewidth]{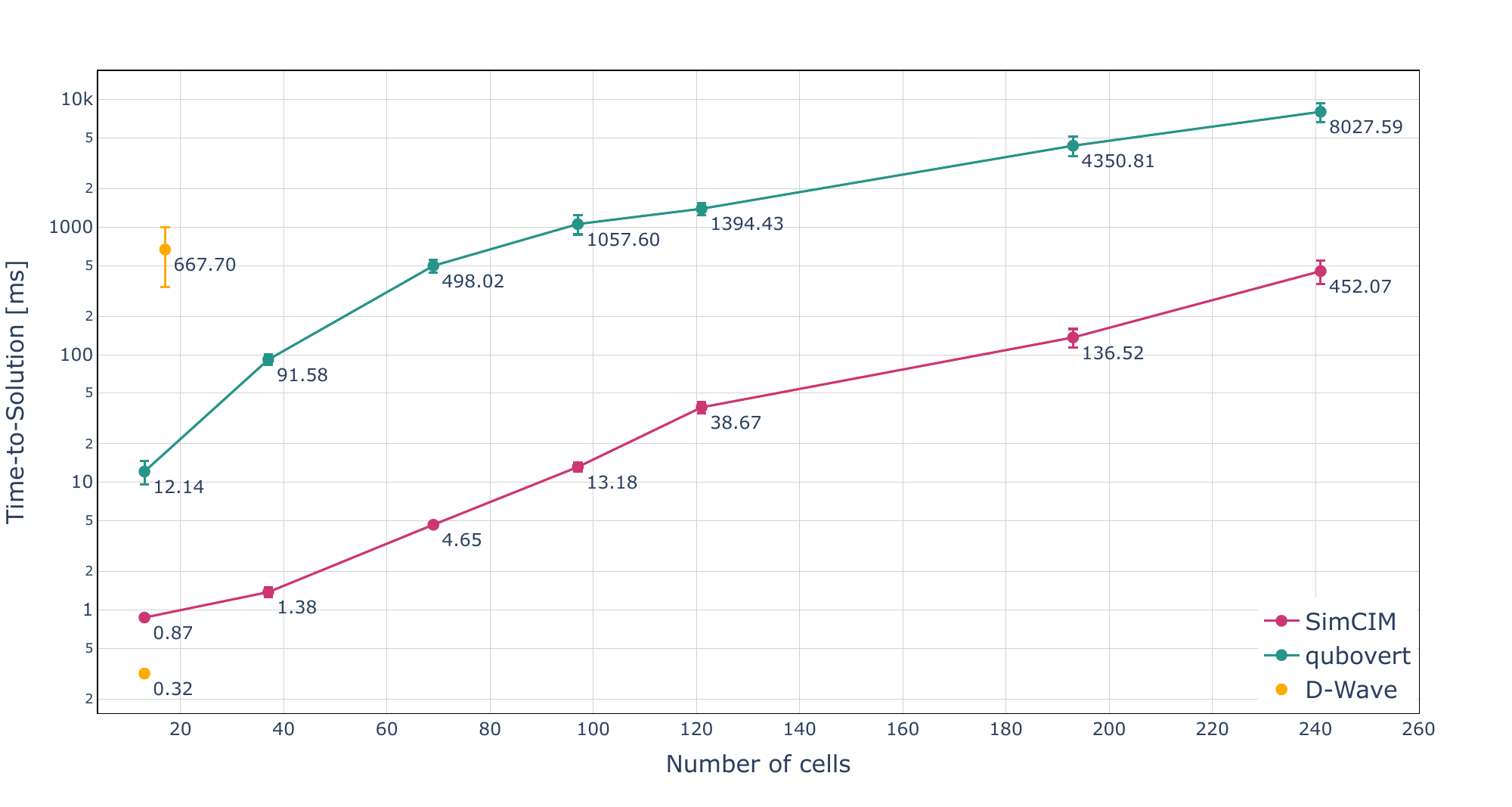}
    \vskip -3mm
    \caption{Time to the feasible solution (ms) depends on cores size for D-Wave quantum computer, SimCIM and qubovert's implementation of simulated annealing.}
    \label{fig:feasible_tts}
\end{figure}

We run D-Wave in a pure quantum sampler mode using the Advantage chip. 
The chip consists of more than 5000 qubits with 15-link connectivity. 
Due to the limited connectivity of physical spins, it is necessary to introduce an additional mapping of the Ising model corresponding to the problem being solved to the physical graph of the D-Wave device. 
We did this using D-Wave Leap SDK embedding tools. 
SimCIM and classical solver were run on Xeon E-2288G CPU 3.70GHz 128 GB RAM 512 GB HDD Geforce GTX 1080ti.
The benchmark results are shown in  Fig.~\ref{fig:feasible_tts}. 

It is worth noting that in addition to the core structure, the number of once- and twice-burnt fuel elements directly affects the space of feasible solutions. 
We  investigated the search space of the 193-cell core depending on the number of amount of burnt cells. 
Results can be viewed in Appendix in Fig.~\ref{fig:feasible_conf}.

\begin{figure}[!h]
\centering
\includegraphics[width=1\linewidth]{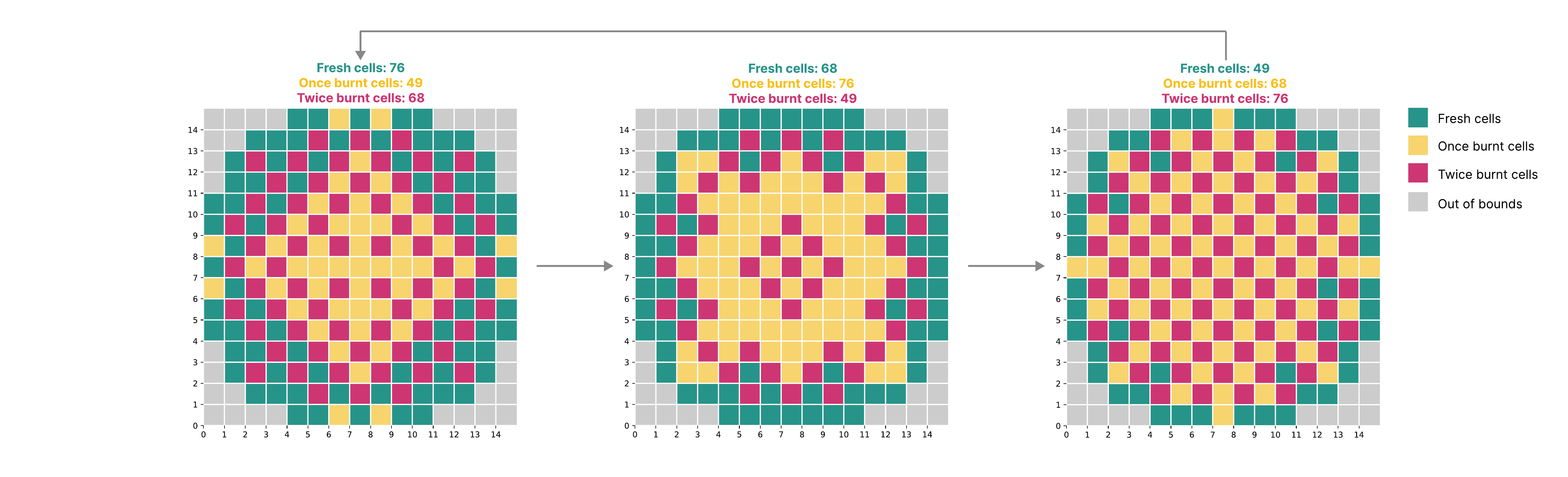}
\vskip -3mm
\caption{Full reloading cycle for 193 cells 4-loop Westinghouse pressurized water reactor (PWR). }
\label{fig:reload_cycle}
\end{figure}

Further research was aimed at finding cycled reloading patterns. 
Cycled reloading patterns are precomputed once during the design of nuclear power plant reactors and could be used as a reloading pattern template for core operation. 
Using results for different once and twice-burnt cells configurations, we found 4 unique feasible full reloading cycles, one of which is displayed in Fig.~\ref{fig:reload_cycle}. 
Each configuration meets all initial constraints, but it is accepted that they can be asymmetrical due to the characteristics of 193 cell core configuration. Every found configuration is shown in Appendix.

\section{Discussion}

In this paper, we have presented a comprehensive formulation of the in-core reloading pattern optimization problem for nuclear reactors that is compatible with quantum annealing devices.
We have used the concept of prohibitive rules and demonstrated how these rules can be translated into the QUBO form. 
We developed a flexible QUBO formulation of the problem with the ability to include additional constraints (such as rotational and mirror symmetry).

We have demonstrated a proof-of-concept solution to the reloading pattern optimization problem using a quantum device and quantum-inspired algorithm 
The map of feasible patterns on the Westinghouse pressurized water reactor (PWR) according to heuristic rules~\cite{Galperin1999} has been found.
That map has been used to analyze patterns in the feasibility of configurations, and we found 4 unique fully feasible closed reloading cycle patterns.

Further research may be aiming to increase the effectiveness of the model. 
Also, additional physical constraints may be added to the QUBO model, such as adding real reactivity parameters and calculating the flux of the configuration, 
so the problem of reloading is solved at the core design process.

One of the distinct goals of the community of researchers in quantum technologies is to find practical applications of quantum computing algorithms for industry-relevant problems, and one of the approaches has been shown here. 
At the current state, we still have not reached a practical advantage in the field of optimization, but it is widely believed that quantum technologies and algorithms have the potential to be more effective than their classical counterparts.
While quantum devices are still in the early development state, quantum algorithms already may be tested and used for real-life problems of limited scale. 
As we also expect, quantum-inspired and hybrid quantum-classical algorithms will find various applications in the near future.

\section*{Acknowledgments} 

This research has been supported by the Russian Quantum Center (QBoard).
We acknowledge online cloud access to the quantum annealing device produced by D-Wave Systems (all the data for this study have been collected during the availability of the device).
S.R. Usmanov, A.A. Bozhedarov,  E.O. Kiktenko, and A.K. Fedorov also thank the support of the RSF grant (19-71-10092).

\appendix

\section{Symmetry constraints}
\begin{figure}[!h]
     \centering
     \begin{subfigure}[b]{0.40\textwidth}
         \centering
         \includegraphics[width=\textwidth]{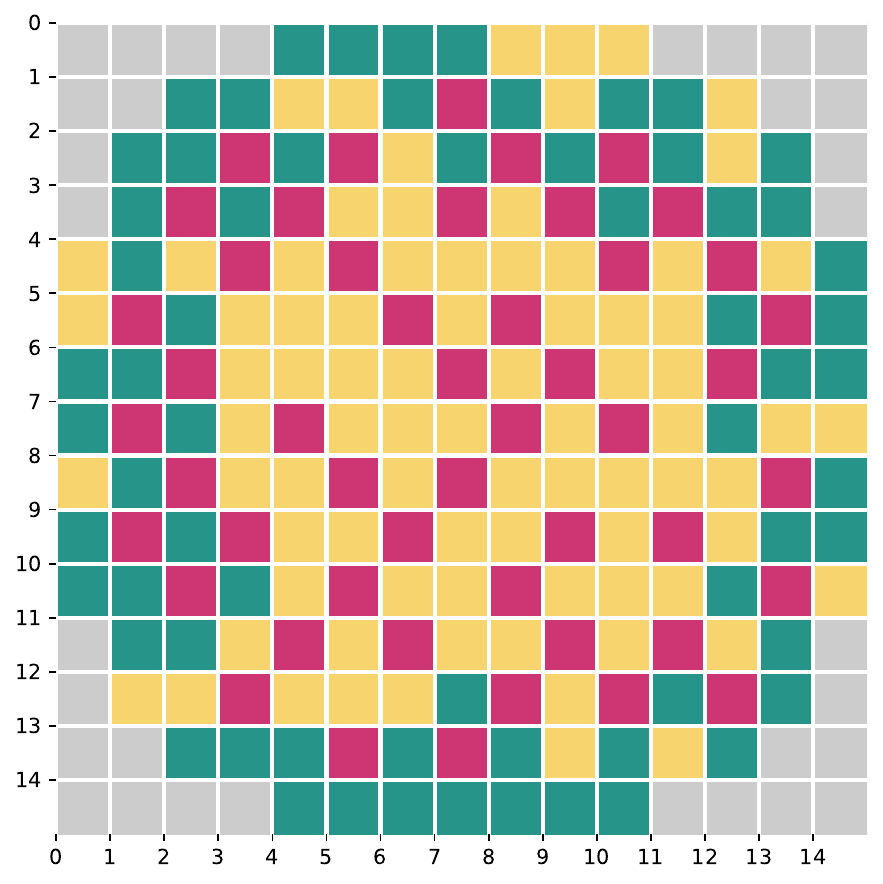}
         \caption{}
         \label{fig:none}
     \end{subfigure}
     \hfill
     \begin{subfigure}[b]{0.40\textwidth}
         \centering
         \includegraphics[width=\textwidth]{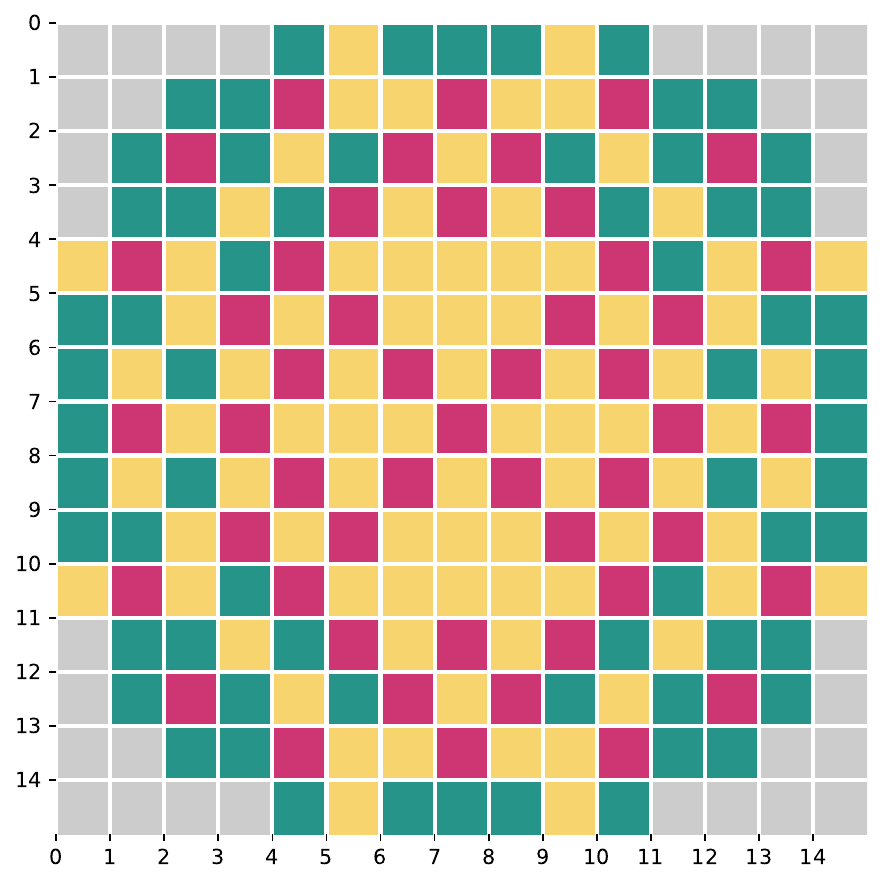}
         \caption{}
         \label{fig:mir}
     \end{subfigure}
     \hfill
     \begin{subfigure}[b]{0.40\textwidth}
         \centering
         \includegraphics[width=\textwidth]{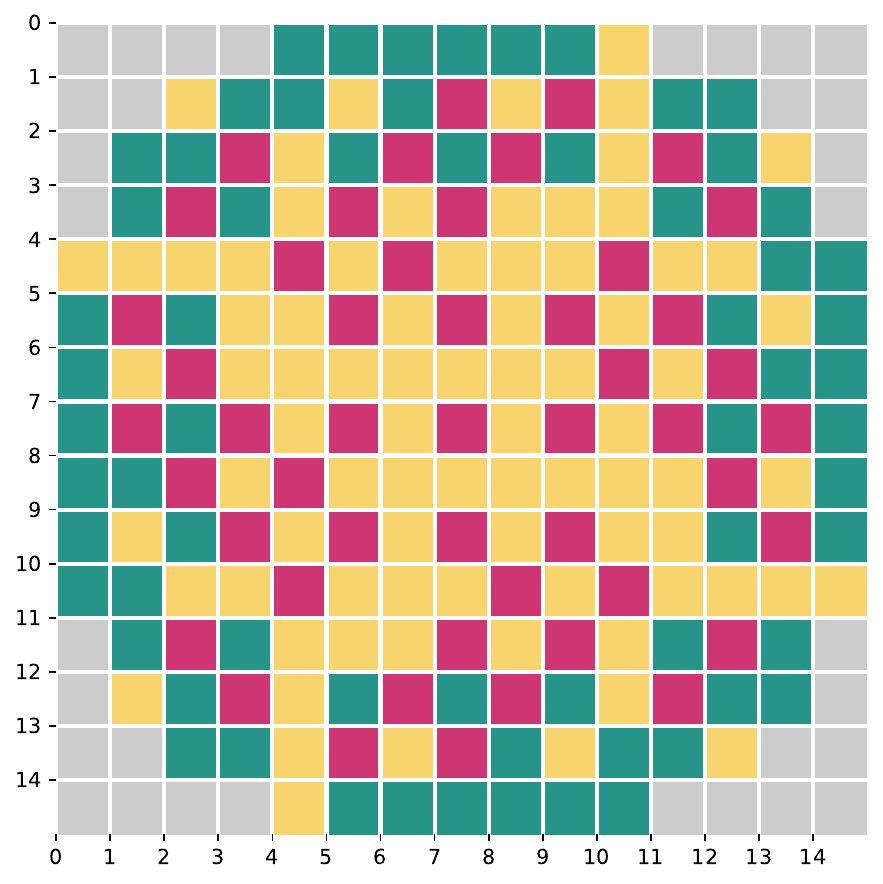}
         \caption{}
         \label{fig:rot}
     \end{subfigure}   
    \hfill
     \begin{subfigure}[b]{0.40\textwidth}
         \centering
         \includegraphics[width=\textwidth]{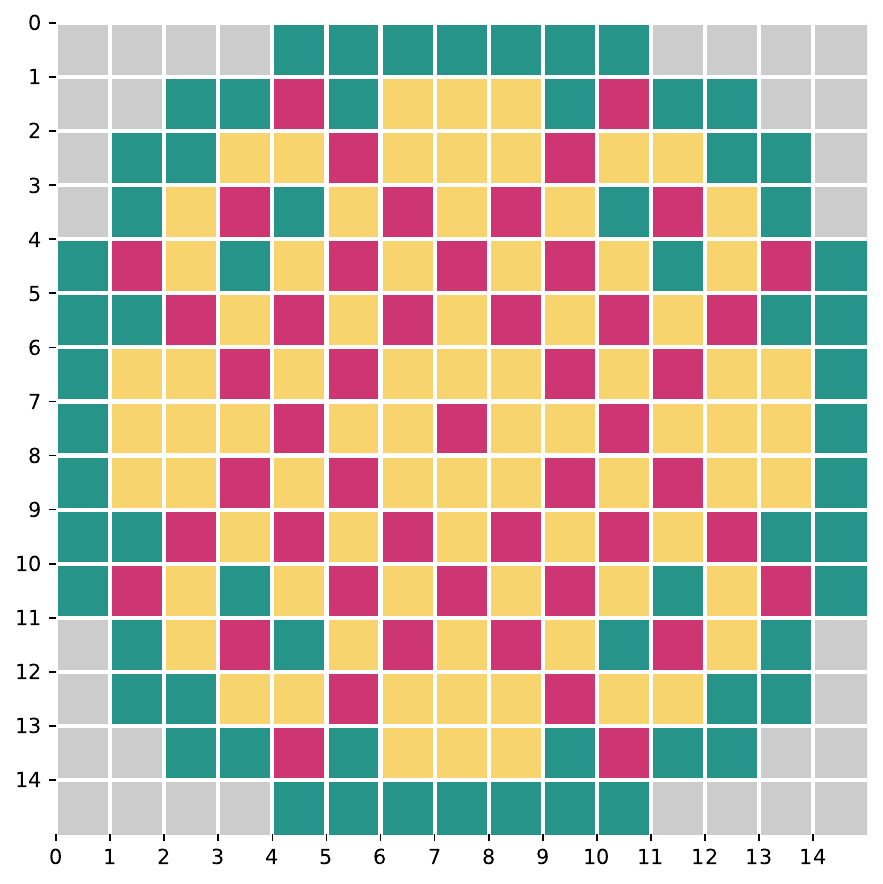}
         \caption{}
         \label{fig:all}
     \end{subfigure}
        \caption{Configuration solutions obtained with the use of SimCIM: 
        without any symmetry constraints (a), with mirror symmetry (b), with rotational symmetry (c), and with mirror and rotational symmetry constraints simultaneously 
        (64 fresh, 80 once burnt, and 49 twice burnt elements) for a 4-loop Westinghouse pressurized water reactor.}
        \label{fig:symms}
\end{figure}

Initially, symmetry constraints are not included in the original formulation, but since the resulting configurations, usually are symmetrical~\cite{Lindley2014, Meneses2010, Dehart1995} we added them. 
We run the experiment with 4 different constraint options:
\begin{enumerate}
	\item all symmetry constraints have been disabled;
	\item mirror symmetry constraints have been enabled;
	\item rotational symmetry constraints have been enabled;
	\item both rotational and mirror symmetry constraints have been enabled.
\end{enumerate}
All the other constraints were enabled for every experiment.
The obtained configurations are shown in Fig.~\ref{fig:symms}. 
Results demonstrate an influence of symmetry constraints on the feasible core configuration.

\section{Feasible core configurations}

\begin{figure}[!h]
\centering
\includegraphics[width=0.8\linewidth]{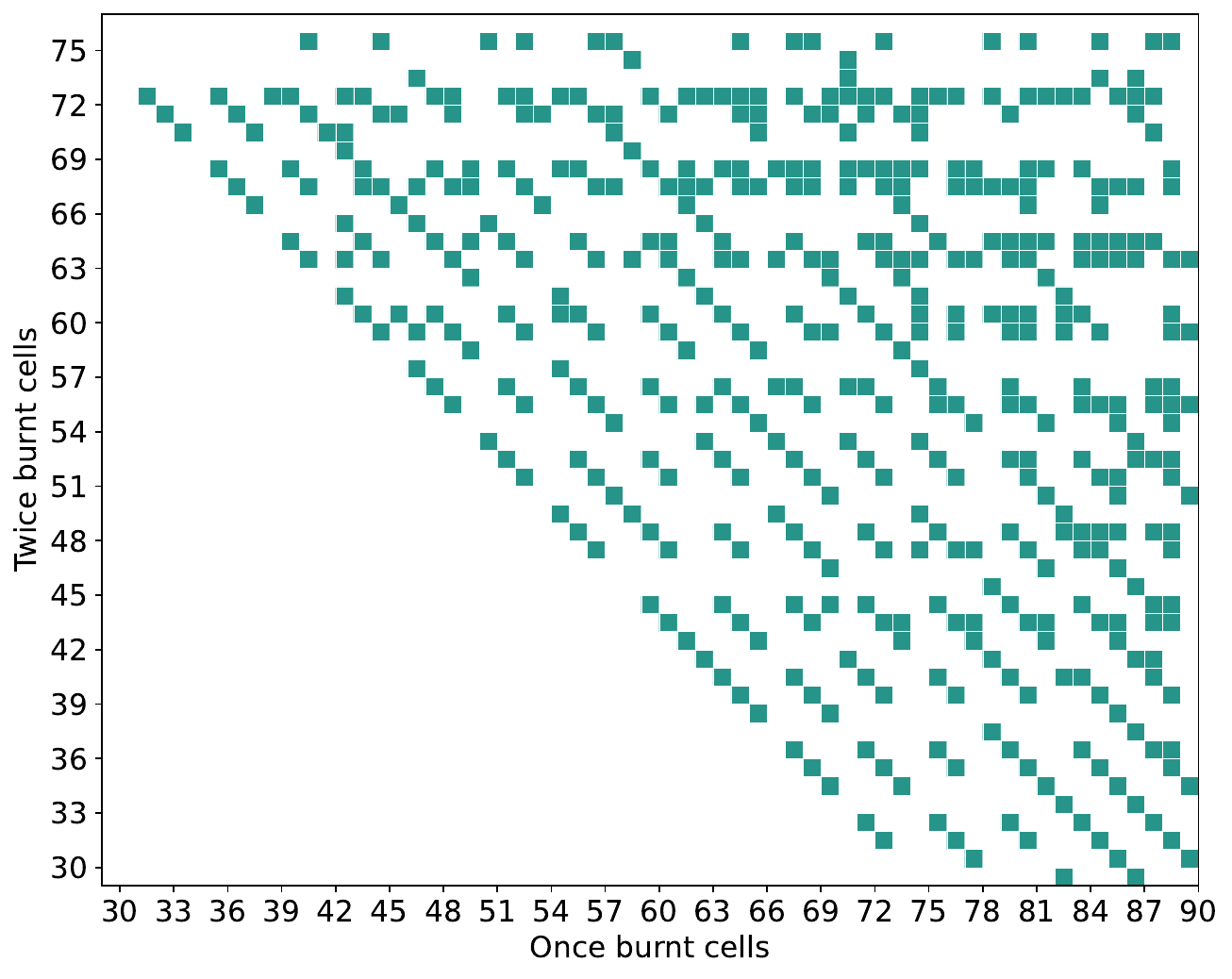}
\vskip -3mm
\caption{Feasible core configurations map for various once and twice-burnt fuel elements for a 193 cells 4-loop Westinghouse pressurized water reactor (PWR).}
\label{fig:feasible_conf}
\end{figure}

Not each core configuration can be achieved without violations of constraints. 
The number of different types of cells directly affects the space of feasible solutions. 
We run the experiment on the 193-cell core configurations of once and twice-burnt cells: each was in the range between and including 30 and 90. 
The result is shown in Fig.~\ref{fig:feasible_conf}. 
Each green square corresponds to a feasible solution for some number of once and twice-burned cells in configurations (74 cells). 
Firstly, not every configuration can be feasible. Secondly, a distinct feature is a clear diagonal border showing that the configuration should have a minimal number of once and twice-burned cells to be feasible. 
Also, there is an upper bound (there should not be more than 75 twice burnt cells in the configuration).

\section{Cycled reloading patterns}
Cycled reloading patterns
are precomputed once during the design of nuclear power plant reactors and could be used as a reloading pattern template for core operation. Using previous results and special algorithm we found 12 cycles with permutations or 4 cycles without them. These 4 cycles in format (FRESH, ONCE BURNT, TWICE BURNT) are:
\begin{subequations}
 \begin{align*}
    &(76, 49, 68) \longrightarrow (68, 76, 49) \longrightarrow (49, 68, 76) \\
    &(72, 49, 72) \longrightarrow (72, 72, 49) \longrightarrow (49, 72, 72) \\
    &(68, 57, 68) \longrightarrow (68, 68, 57) \longrightarrow (57, 68, 68) \\
    &(65, 64, 64) \longrightarrow (64, 65, 64) \longrightarrow (65, 64, 65)
 \end{align*}
\end{subequations}

\section*{References}
\bibliographystyle{iopart-num}
\bibliography{bibliography.bib}

\providecommand{\newblock}{}
\begin{thebibliography}{10}
\expandafter\ifx\csname url\endcsname\relax
  \def\url#1{{\tt #1}}\fi
\expandafter\ifx\csname urlprefix\endcsname\relax\def\urlprefix{URL }\fi
\providecommand{\eprint}[2][]{\url{#2}}

\bibitem{Yarkoni2022}
Yarkoni S, Raponi E, Bäck T and Schmitt S 2022 {\em Reports on Progress in
  Physics\/} {\bf 85}

\bibitem{Fedorov2021}
Boev A~S, Rakitko A~S, Usmanov S~R, Kobzeva A~N, Popov I~V, Ilinsky V~V,
  Kiktenko E~O and Fedorov A~K 2021 {\em Scientific Reports\/} {\bf 11} 13183

\bibitem{Boev2023}
Boev A, Usmanov S, Semenov A, Ushakova M, Salahov V, Mastiukova A, Kiktenko E
  and Fedorov A 2023 {\em Frontiers in Physics\/} {\bf 10}

\bibitem{Fedorov2022}
Fedorov A~K, Gisin N, Beloussov S~M and Lvovsky A~I 2022 Quantum computing at
  the quantum advantage threshold: a down-to-business review

\bibitem{Orus2019}
Or{\'u}s R, Mugel S and Lizaso E 2019 {\em Reviews in Physics\/} {\bf 4} 100028
  ISSN 2405-4283

\bibitem{DoBinh2014}
Do B, Huy N and Hai N 2014 {\em Kerntechnik\/} {\bf 79} 511--517

\bibitem{Meneses2018ArtificialIM}
Meneses A, de~Lima A~M~M and Schirru R 2018 Artificial intelligence methods
  applied to the in-core fuel management optimization

\bibitem{LI2022111950}
Li Z, Wang J and Ding M 2022 {\em Nuclear Engineering and Design\/} {\bf 397}
  111950 ISSN 0029-5493

\bibitem{Whyte2020}
Whyte A 2020 {\em Surrogate Model Optimisation for PWR Fuel Management\/} Ph.D.
  thesis

\bibitem{Galperin1999}
Galperin A 1995 {\em Nuclear Science and Engineering\/} {\bf 119} 144--152

\bibitem{Horelik2013}
Horelik N, Herman B, Forget B and Smith K 2013 {\em American Nuclear Society -
  ANS\/} ISSN 978-0-89448-700-2

\bibitem{Lindley2014}
Lindley B, Fiorina C, Gregg R, Franceschini F and Parks G 2014 {\em Progress in
  Nuclear Energy\/} {\bf 85}

\bibitem{Meneses2010}
Meneses A, De~Lima A and Schirru R 2010 {\em Artificial Intelligence Methods
  Applied to the In-Core Nuclear Fuel Management Optimization\/} ISBN
  978-953-307-110-7

\bibitem{Dehart1995}
Dehart M and Bowman S 1995

\bibitem{Glover2019}
Glover F, Kochenberger G and Du Y 2019 {\em 4OR\/} {\bf 17}

\bibitem{Yamamoto2017}
Yamamoto Y, Aihara K, Leleu T, Kawarabayashi K~i, Kako S, Fejer M, Inoue K and
  Takesue H 2017 {\em npj Quantum Information\/} {\bf 3} 49

\bibitem{Tiunov2019}
Tiunov E~S, Ulanov A~E and Lvovsky A~I 2019 {\em Opt. Express\/} {\bf 27}
  10288--10295

\bibitem{Sabundjian2012}
Gaiane S, Delvonei A, Belchior A, Marcelo R, Conti N, Walmir T, Macedo A, Pedro
  U, de~Mesquita~Roberto, Masotti F and de C 2012 The behavior of angra 2
  nuclear power plant core for a small break loca simulated with relap5 code
  vol 1529

\bibitem{Freixa2007}
Freixa J 2007 {\em SBLOCA with Boron Dilution in Pressuirzed Water Reactors,
  Impact to the Operation and Safety\/} Ph.D. thesis

\bibitem{INGERSOLL201484}
Ingersoll D, Houghton Z, Bromm R and Desportes C 2014 {\em Desalination\/} {\bf
  340} 84--93 ISSN 0011-9164

\bibitem{Aramon2019}
Aramon M, Rosenberg G, Valiante E, Miyazawa T, Tamura H and Katzgraber H~G 2019
  {\em Frontiers in Physics\/} {\bf 7} ISSN 2296-424X

\bibitem{Stollenwerk2020}
Stollenwerk T, Michaud V, Lobe E, Picard M, Basermann A and Botter T 2020 Image
  acquisition planning for earth observation satellites with a quantum annealer

\end{thebibliography}

\end{document}